\documentclass[10pt, a4paper]{article}

\long\def\symbolfootnote[#1]#2{\begingroup%
\def\thefootnote{\fnsymbol{footnote}}\footnote[#1]{#2}\endgroup} 
\def\araa{ARA\&A}
\def\apj{ApJ}
\def\apjl{ApJ}
\def\aap{A\&A}
\def\mnras{MNRAS}
\def\nat{Nature}
\def\procspie{Proc.~SPIE}

\usepackage{lineno,graphicx,color,authblk,amsmath,amssymb,setspace,soul,hyperref,caption}
\usepackage[a4paper, margin=0.8in]{geometry}

\newcommand{\nustar}{\textit{NuSTAR}}

\newcommand{\xmm}{{\it XMM-Newton}}

\author[1]{Michael L. Parker}
\author[1]{Ciro Pinto}
\author[1]{Andrew C. Fabian}
\author[1]{Anne Lohfink}
\author[1]{Douglas J. K. Buisson}
\author[1]{William Alston}
\author[2]{Erin Kara}

\author[3]{Edward M. Cackett}
\author[3]{Chia-Ying Chiang}
\author[4]{Thomas Dauser}
\author[5]{Barbara De Marco}
\author[6]{Luigi C. Gallo}
\author[7]{Javier Garcia}
\author[7]{Fiona A. Harrison}
\author[8]{Ashley L. King}
\author[9]{Matthew J. Middleton}
\author[10]{Jon M. Miller}
\author[11]{Giovanni Miniutti}
\author[2]{Christopher S. Reynolds}
\author[12]{Phil Uttley}
\author[1]{Ranjan Vasudevan}
\author[1]{Dominic J. Walton}
\author[8]{Daniel R. Wilkins}
\author[10]{Abderahmen Zoghbi}

\affil[1]{\small Institute of Astronomy, Madingley Road, Cambridge, CB3 0HA, UK}
\affil[2]{Department of Astronomy, University of Maryland, College Park, MD 20742-2421, USA}
\affil[3]{Department of Physics and Astronomy, Wayne State University, Detroit, MI 48201, USA}
\affil[4]{Remeis Observatory and ECAP, Universitat Erlangen-Nurnberg, Sternwartstr. 7, D-96049, Bamberg, Germany}
\affil[5]{Max-Planck-Institut fur extraterrestrische Physik, Giessenbachstrasse, D-85748 Garching, Germany}
\affil[6]{Department of Astronomy and Physics, Saint Mary's University, 923 Robie Street, Halifax, Nova Scotia B3H 3C3, Canada}
\affil[7]{Space Radiation Laboratory, California Institute of Technology, 1200 E California Blvd, MC 249-17, Pasadena, CA 91125, USA}
\affil[8]{Kavli Institute for Particle Astrophysics and Cosmology, Stanford University, 452 Lomita Mall, Stanford, CA 94305, USA}
\affil[9]{School of Physics and Astronomy, University of Southampton, Southampton, SO17 1BJ, UK}
\affil[10]{Department of Astronomy, University of Michigan, 1085 South University Avenue, Ann Arbor, MI 48109, USA}
\affil[11]{Centro de Astrobiologia (CSIC-INTA), Dep. de Astrofísica, ESAC, PO Box 78, 28691 Villanueva de la Cañada, Madrid, Spain}
\affil[12]{Astronmical Institute ‘Anton Pannekoek’, University of Amsterdam, Postbus 94249, NL-1090 GE Amsterdam, the Netherlands}

\title{The response of relativistic outflowing gas to the inner accretion disk of a black hole}

\begin{document}
\maketitle

\textbf{Active galactic nucleus (AGN) feedback is the process by which supermassive black holes in the centres of galaxies may moderate the growth of their hosts\cite{Silk98}. 
Gas outflows from supermassive black holes release huge quantities of energy into the interstellar medium\cite{Fabian12}, clearing the surrounding gas. The most extreme of these, the ultra-fast outflows (UFOs), are the subset of X-ray detected outflows with velocities higher than 10,000~km~s$^{-1}$, believed to originate in relativistic disc winds, a few hundred gravitational radii from the black hole\cite{Nardini15}. The absorption features produced by these outflows are variable\cite{Cappi09}, but no clear link has been found between the behaviour of the X-ray continuum and the energy or equivalent width of the outflow features due to the long time-scales of quasar variability. Here, we present the detection of multiple absorption lines from an extreme ultra-fast gas flow in the X-ray spectrum of the active galactic nucleus IRAS~13224--3809, at $0.236\pm0.006$ times the speed of light (71,000 km~s$^{-1}$), where the absorption is strongly anti-correlated with the emission from the inner regions of the accretion disk. If the gas flow is identified as a genuine outflow then it is in the fastest 5\% of such winds, and its variability is hundreds of times faster than in other variable winds, allowing us to observe in hours what would take months in a quasar. We find signatures of the wind simultaneously in both low and high energy detectors, which are consistent with a single ionized outflow, linking the two phenomena. The detection of the wind responding to the emission from the inner disk demonstrates a connection between accretion processes occurring on very different scales, with the X-rays from within a few gravitational radii of the black hole ionizing the relativistically outflowing gas as the flux rises.
}

\begin{figure}[t]
\centering
\includegraphics[width=16cm]{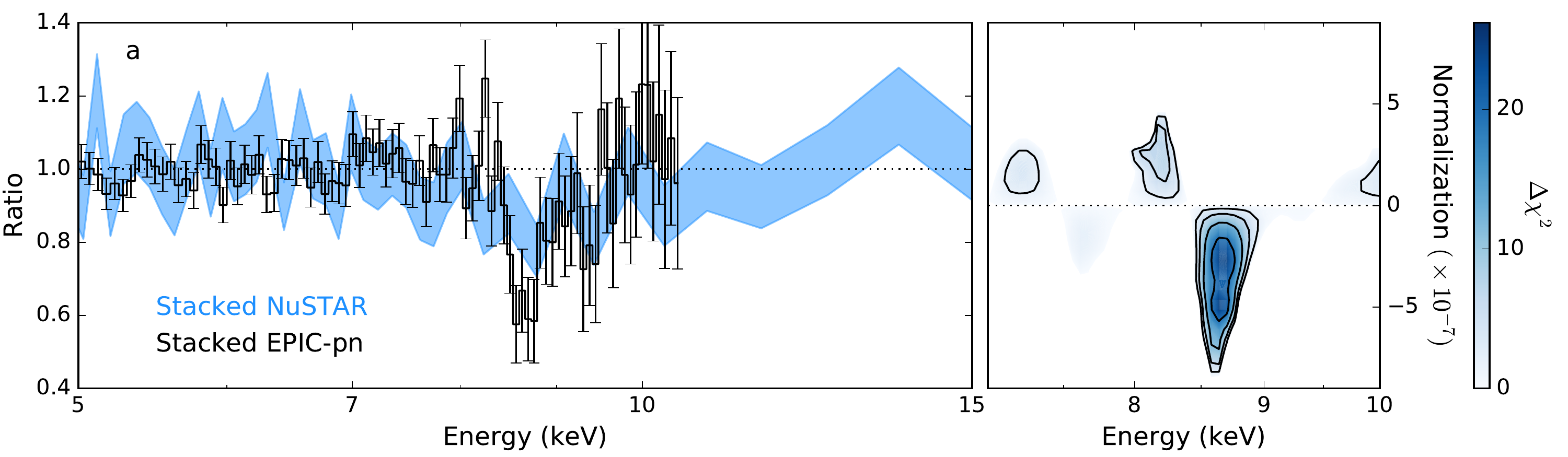}
\includegraphics[width=12cm]{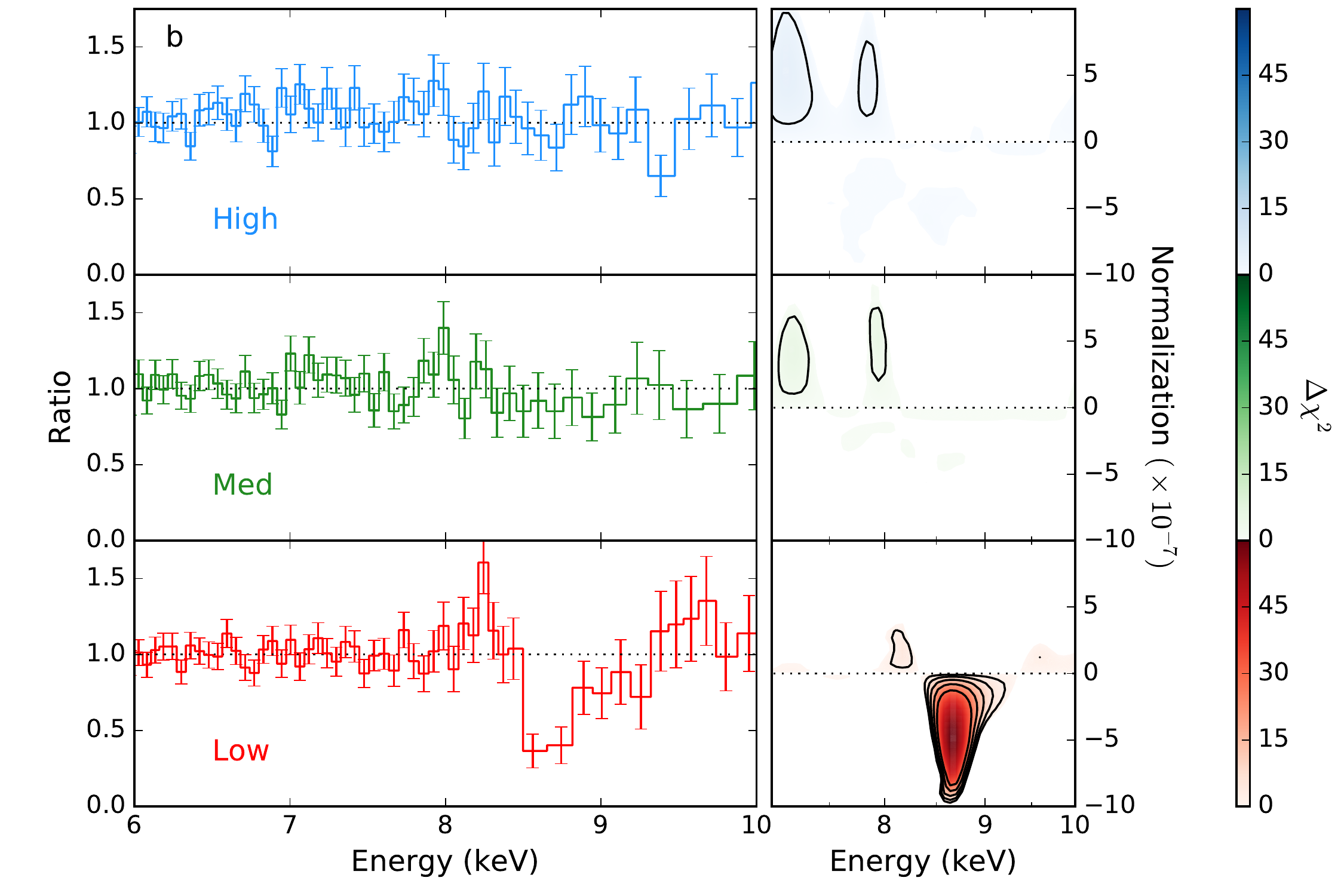}
\caption{\emph{a:} High energy \xmm\ EPIC-pn and \nustar\ residuals when fit with a relativistic reflection model. A strong absorption feature is visible at $\sim8.6$~keV in the EPIC-pn data, and is consistent with the lower-resolution \nustar\ spectrum. The right panel shows the results of a line search over this region, where the significance peaks at $>4\sigma$. \emph{b:} Residuals of stacked spectra of IRAS~13224-3809 at three different flux levels, fit with a relativistic reflection model, showing the flux dependence of the outflow, and line scan results for each of the three spectra. There is no significant absorption present in the medium or high flux spectra, while in the low state the line is present at $\sim7\sigma$. All energies are in the rest frame of the source. Error bars are $1\sigma$.}
\label{fig_fluxresolved}
\end{figure}

IRAS~13224--3809 is a low-redshift ($z=0.0658$) NLS1. These galaxies are characterised by having low mass, high accretion rate black holes\cite{Mathur00}, some of which have evidence for relativistic outflows \cite{Longinotti15}. IRAS~13224--3809 is the most variable AGN in the X-ray band\cite{Ponti12}, changing by up to two orders of magnitude in flux in only a few hours. This source was the target of a recent 1.5~Ms continuous observing campaign with the X-ray Multi-Mirror Mission (\xmm \cite{Jansen01}) and 500~ks with \nustar \cite{Harrison13} intended to study the rapid variability, from minutes to weeks. IRAS~13224--3809 is well known as a source with strong relativistic reflection from the inner disk \cite{Fabian13}, where fluorescent emission, produced by X-ray illumination of the inner accretion disk, is blurred and skewed by the strong relativistic effects close to the event horizon. This produces a characteristic broad iron line at 6--7~keV and a Compton scattered `hump' at high energies. In this case, reflection spectroscopy revealed near maximal spin, and a spectrum dominated by reflection, implying a high degree of light-bending close to the event horizon. The iron K and L lines can also be modelled as a relavitistic absorption edge \cite{Leighly97,Boller03}, however the measurement of X-ray reverberation, which measures the delay between the continuum and reflected emission\cite{Kara13}, demonstrated that the line is produced by reflection and that we have an unobstructed view of the inner accretion disk.

The source is spectrally very soft, hence the very high energy band ($>8$~keV) has not been well studied previously, due to a shortage of photons in any given observation. By combining all the data from the 2016 observing campaign into a single stacked spectrum and constraining the continuum flux using \nustar , we are able to reveal the presence of a structured absorption feature at $\sim8.6$~keV in the rest frame of the source, significant at $>4\sigma$ (Fig.~\ref{fig_fluxresolved}, top), along with a possible secondary line at $\sim9.2$~keV. The 8.6~keV line is also present, at $>3\sigma$, in the stacked spectrum of the archival data (Fig.~\ref{fig_archival_comparison}), meaning that it must have been present for at least 5 years. Including the absorption line does not affect the results of reflection fitting, the parameters returned are entirely consistent with those found by previous authors\cite{Fabian13,Chiang15} (near maximal spin, inclination of $\sim60$ degrees, and a steep emissivity profile).

Additional evidence for the presence of a UFO is given by the simultaneous detection of three more blueshifted absorption lines in the high spectral resolution Reflection Grating Spectrometer (RGS\cite{denHerder01}) data, which is capable of resolving narrow features at low energies. These features have a combined significance of $>5\sigma$, and are found at the same blueshift as the 8.6~keV line. Three broad features are detected at $\sim$9.5, 13.0, and 16 \AA\ in the low-flux spectrum. They have less significance or are undetected in the higher flux spectra with possible evidence of slight velocity change. Their wavelengths match with the strongest lines predicted by the photoionized UFO model in the RGS energy band: 10.0 $\pm$ 0.5 \AA\ (Ne X + Fe XVIII-XXII blend), 13.2 $\pm$ 0.5 \AA\ (O VIII K$\beta$ + Fe XVIII) and 15.8 $\pm$ 0.5 (O VIII K$\alpha$).

\begin{figure}[t]
\centering
\includegraphics[width=14cm]{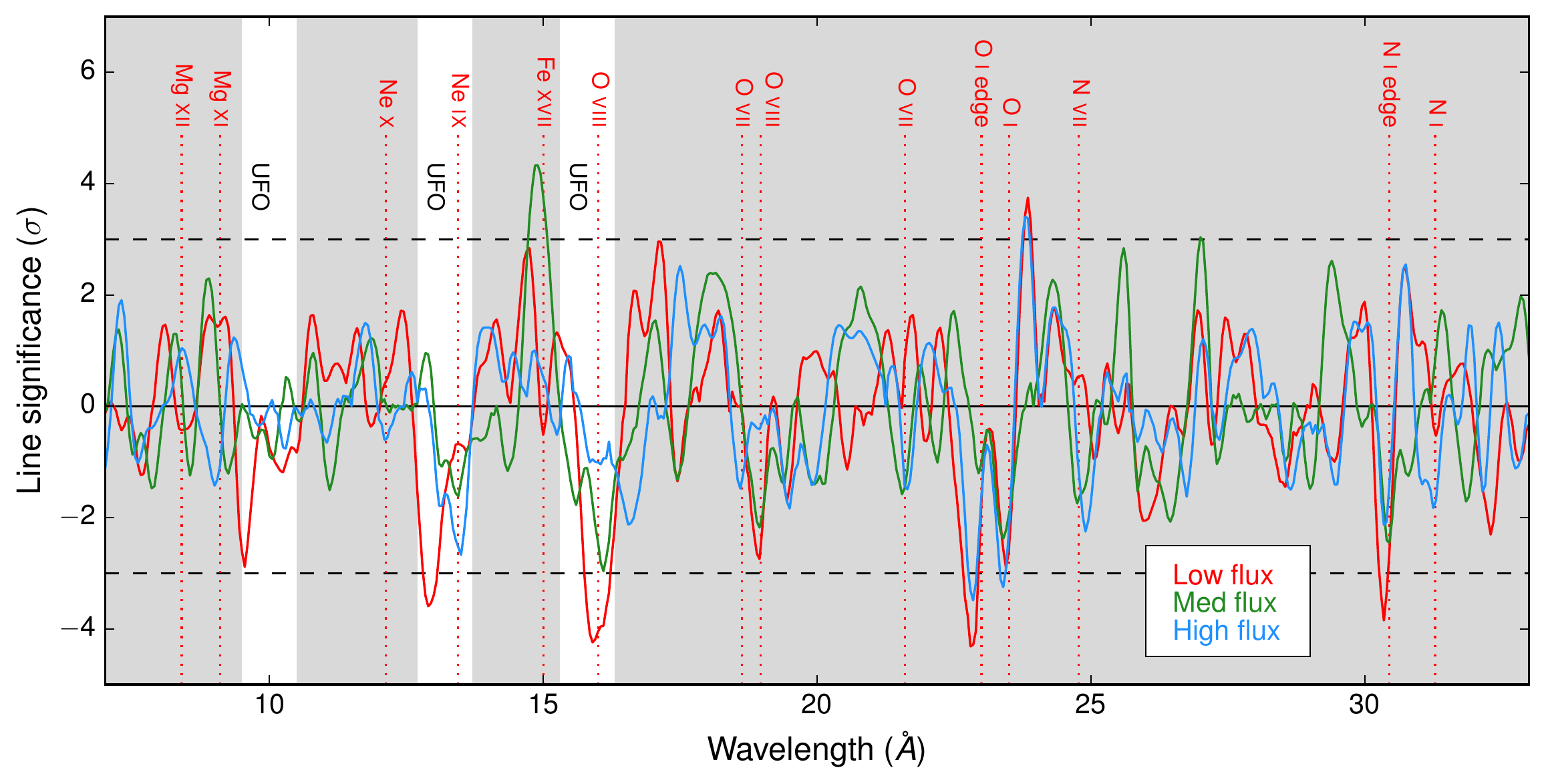}
\caption{Shown is the line significance obtained by a Gaussian fitting the 7-27 \AA\ wavelength range with increments of 0.05 \AA\ and negative values indicating absorption lines. The solid lines indicate the line significance obtained with 1000 km~s$^{-1}$ widths for the three flux-selected RGS spectra. The low-flux RGS spectrum is plotted in red, the intermediate-flux spectrum is in green, and high-flux spectrum in blue. The rest-frame transitions of some relevant interstellar absorption lines are labelled. We highlight the regions where UFO absorption is predicted according to the spectral modelling of the EPIC data (velocity shift of 0.24$c$). The spectral continuum used here is a spline corrected by redshifted and galactic neutral absorption. Wavelengths are in the observed frame.
}
\label{fig_rgs_significance}
\end{figure}

Depending on the ionization of the source, the 8.6~keV feature could be due to either Fe XXV or Fe XXVI, with corresponding outflow velocities of 0.24$c$ or $0.21c$. However, a joint fit to the EPIC-pn and RGS data with a single absorber gives a best fit velocity of $0.236\pm0.006c$, and robustly prefers the faster Fe \textsc{xxv} solution.
The typical outflow velocity of radio quiet AGN is $\sim0.1c$ \cite{Tombesi10}, this line identification puts this outflow in the fastest $\sim$5\% of sources.
The power output is also substantial. Assuming a solid angle $\Omega=2\pi$, the kinetic power of the wind is $\sim4$\% of the Eddington luminosity (see methods), comparable to the 15\% found in the quasar PDS~456\cite{Nardini15}, which is 2--3 orders of magnitude higher in mass ($\sim10^9M_\odot$, compared to $\sim6\times10^6M_\odot$ for IRAS~13224-3809). This indicates that the wind in IRAS~13224-3809 may be capable of driving feedback.
Similarly, the mass outflow rate is 0.4~$\dot{M}_\mathrm{Edd}$, implying that a large fraction of the mass is lost in the wind and that the accretion rate outside the wind launching radius may be super-Eddington. This is very similar to the wind found in the ultraluminous X-ray source NGC~1313~X-1\cite{Pinto16, Walton16}, which also responds to changes in the continuum\cite{Middleton15}, implying that super-Eddington relativistic winds may be a common phenomenon in high accretion rate black holes\cite{Shakura73, King03, Poutanen07}.


IRAS~13224-3809 varies dramatically in flux on short time scales, giving us a unique opportunity to probe the fast variability of the wind. We split the large dataset up into three different flux levels such that the total number of counts in each of the three spectra is equal (Fig.~\ref{fig_fluxresolved}). In the lowest flux state, the line is extremely significant ($7\sigma$), with an equivalent width of 0.24~keV. However, when the flux is higher the line disappears completely, and is not required in the medium and high flux spectra.
This variability of the 8.6~keV line strength must be occurring on relatively short timescales ($<10$ks), as the source variability is so extreme that any delay in the absorption responding to the continuum flux would remove the observed correlation between the two. The average time between the end of a low flux interval and the start of a high flux interval is 4--5~ks, so the line must be responding within this time. Spectrally, the changes in the equivalent width can be explained by changes in either the column density or the ionization of the absorber. However, a causal relationship between the flux of the source and the ionization is expected, and both the flux and the model ionization increase by a factor of $\sim4$ between the low and high flux spectra, consistent with the ionization interpretation. Because of the strong relativistic reflection observed in this source, we can be confident that we are seeing the inner regions of the AGN, and the gas is responding directly to that emission.

\begin{figure}[t]
\centering
\includegraphics[width=12cm]{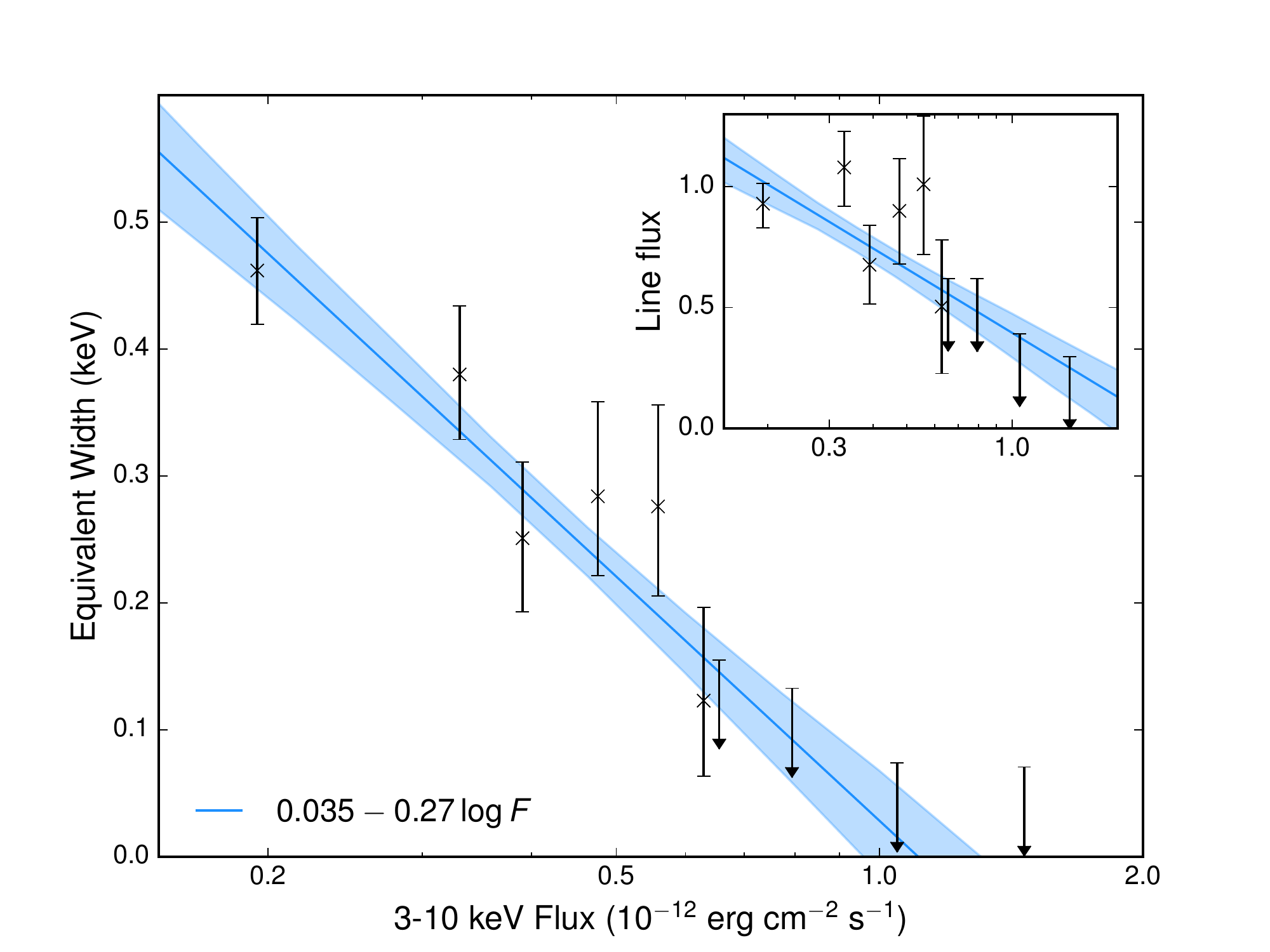}
\caption{ Equivalent width of the EPIC-pn absorption line, plotted against the 3--10~keV flux, for ten flux-resolved spectra. A strong correlation is visible, and is fit with a linear relation (in $\log{F}$), calculated using Bayesian regression (which self-consistently takes into account the upper limits). Shaded regions show the $1\sigma$ confidence regions, calculated by drawing points from the posterior probability distribution. The inset shows the same correlation for the normalized line flux, which demonstrates that the change in equivalent width cannot be explained by a constant flux line. Error bars are $1\sigma$.}
\label{fig_linestrength}
\end{figure}

The strength of the RGS absorption lines also anti-correlates with the flux, and therefore strongly argues in favour of the EPIC and RGS absorbers being part of the same extreme wind. In Fig.~\ref{fig_rgs_significance} we show the significance of the spectral features obtained adopting the (RGS fitted) spline continuum. We computed the confidence level of the three main absorption lines in the low-flux spectrum, where they are significantly detected, as with the EPIC-pn. We obtain 2.1$\sigma$, 2.9$\sigma$, and 3.4$\sigma$ for the 9.5 \AA , 13 \AA , and 16 \AA\ absorption lines, respectively, which, since they have the same velocity shift and broadening, gives a cumulative 5.1$\sigma$ detection.

We investigate this in more detail by further subdividing the data into 10 flux-resolved spectra, again with the same number of counts in each. Fitting these data with the same model and allowing the normalization of the Gaussian line to vary between them with the velocity fixed gives a very strong correlation between the equivalent width of the line and the 3--10~keV flux (Fig.~\ref{fig_linestrength}). The significance of this correlation is $>99.99$\%. This rapid variability clearly demonstrates, for the first time, that the outflow is responding to the source flux.
Variability of outflows has been reported before\cite{Cappi09}, but has never shown a clear correlation with the continuum behaviour. The correlation we find can potentially be explained as the absorption lines disappearing due to the increased ionization when the flux is high. Alternatively, this could be a geometric effect, where the line of sight from the X-ray corona intersects the wind at low source heights and misses it when the source is high or vertically extended, or the line could be produced in an ionized layer on the surface of the inner disk, absorbing the reflected emission only\cite{Gallo11}. Since the reflection fraction scales inversely with flux \cite{Chiang15}, this could produce the observed correlation.

The discovery of a rapidly variable outflow, analogous to a quasar wind, in a low-mass, low-redshift AGN gives us a new window onto AGN feedback. The simultaneous detection of a relativistic outflow in the RGS and EPIC-pn spectra means that we can accurately characterise the physical properties of the wind, demonstrating that both sets of features can be produced by the same physical wind. The rapid variability, in a source where we can see clearly into the inner accretion disk, allows us to measure the wind responding to the X-rays originating a few gravitational radii away from the black hole, linking the physical processes occurring on these different scales. This discovery may allow us to uncover how relativistic winds are launched and accelerated.


\subsection*{Acknowledgements}
M.L.P., C.P., A.C.F., and A.L. acknowledge support from the European Research Council through Advanced Grant on Feedback 340492. 
W.N.A. and G.M. acknowledge support from the European Union Seventh Framework Programme (FP7/2013-2017) under grant agreement n.312789, StrongGravity. 
D.J.K.B. acknowledges support from the Science and Technology Facilities Council (STFC). This work
is based on observations with XMM-Newton, an ESA science mission with instruments and contributions directly funded by ESA Member States and NASA.
D.R.W. is supported by NASA through Einstein Postdoctoral Fellowship grant number PF6-170160, awarded by the Chandra X-ray Center, operated by the Smithsonian Astrophysical Observatory for NASA under contract NAS8-03060.
This work made use of data from the
NuSTAR mission, a project led by the California Institute of
Technology, managed by the Jet Propulsion Laboratory, and
funded by NASA. This research has made use of the NuSTAR
Data Analysis Software (NuSTARDAS) jointly developed by
the ASI Science Data Center and the California Institute of
Technology.

\subsection*{Author contributions}
M.L.P. wrote the manuscript with comments from all authors and performed the flux-resolved EPIC-pn analysis and line detections. C.P. analysed the RGS data and did the physical modelling. A.C.F. lead the \xmm\ proposal. All authors were involved with the proposal at various stages.

\subsection*{Author Information Statement}
Reprints and permissions information is available at www.nature.com/reprints. Correspondence and requests for materials should be addressed to \url{mlparker@ast.cam.ac.uk}. The authors declare no competing financial interests.

\subsection*{Data availability statement}
All data used in this work is publicly available. The \xmm\ observations can be accessed from the \xmm\ science archive (\url{nxsa.esac.esa.int/nxsa-web/}) and the \nustar\ data from the HEASARC archive (\url{heasarc.gsfc.nasa.gov/docs/archive.html}). Figure data is available from the authors.

\setcounter{figure}{0}  

\captionsetup[figure]{name=Extended Data Figure}
\captionsetup[table]{name=Extended Data Table}

\pagebreak

\section*{Methods}
\begin{figure}[h]
\includegraphics[width=\linewidth]{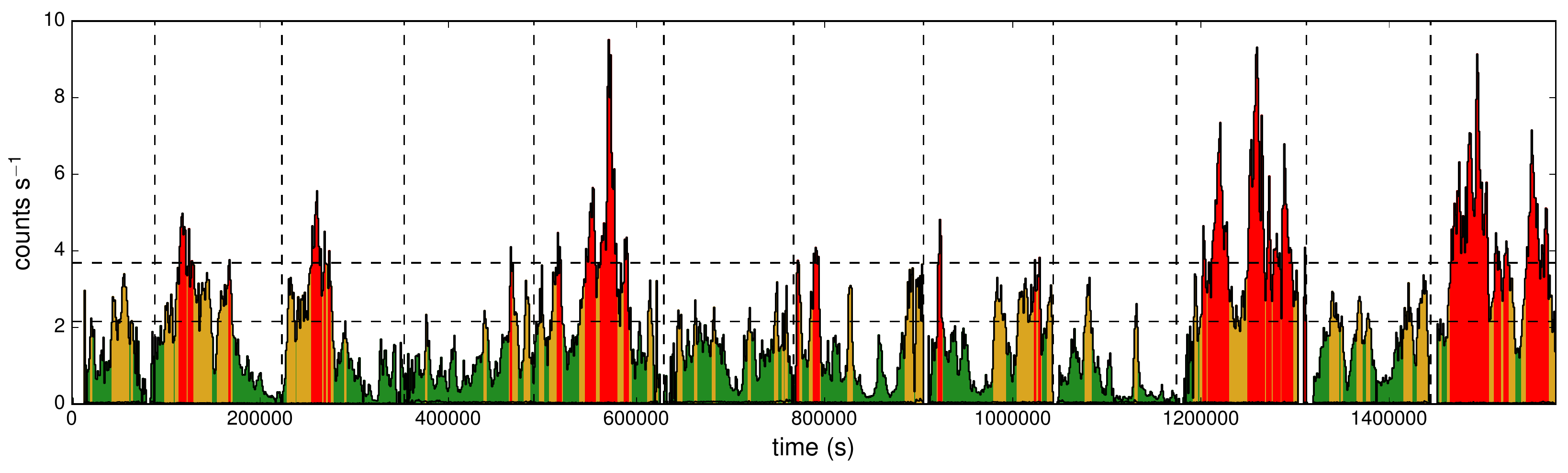}
\caption{0.3--10~keV lightcurve of the new observations of IRAS~13224 in 1~ks bins, with gaps between observations removed and divided into flux levels. The vertical dashed lines show where observations start and finish, and the horizontal lines show the threshold flux levels. Low, medium, and high flux intervals are distributed throughout the lightcurve, and are coloured green, yellow and red, respectively.}
\label{fig_lightcurve}
\end{figure}
\subsection*{Data reduction}

We use all the available \xmm\ data, both from our recent observing campaign (PI Fabian) and from the archive. The EPIC-pn data are reduced using \xmm\ Science Analysis System (SAS) version 15.0.0 \textsc{epproc} tool. The EPIC observations were made in large window mode. We extract source counts from a 30$^{\prime\prime}$ diameter circular region centred on the source coordinates, and background counts from a $\sim60^{\prime\prime}$ circular region nearby on the same chip, avoiding contaminating sources, chip edges, and the region where the internal background due to copper is high,  and filter the data for background flares. We create separate stacked spectra of the archival and new data using the \textsc{addspec} ftool. We extract full band (0.3--10~keV) lightcurves for each spectrum, shown in Extended Data Figure~\ref{fig_lightcurve}, and divide the lightcurve into low, medium, and high flux intervals such that each flux band contains the same total number of counts (thus the low flux intervals are much longer than the high flux intervals). We then extract spectra corresponding to each flux level from each observation, and combine them using \textsc{addspec}. We bin all the EPIC-pn spectra to a signal-to-noise ratio of 6, after background subtraction, and to oversample the spectral resolution by a factor of 3.

The RGS camera consists of two similar detectors, which have high effective area and high spectral resolution between 7 and 38 \AA . The second order spectra cover the 7--18 \AA\ wavelength range and provide double the spectral resolution. We correct for contamination from soft-proton flares following the XMM-SAS standard procedures. For each exposure, we extracted the first- and second-order RGS spectra in a cross-dispersion region of 1$^\prime$ width, centred on IRAS sky coordinates. We have extracted background spectra by selecting photons beyond the 98\% of source point-spread-function. The background spectra were consistent with those from blank field observations. Using the SAS task {\sc rgscombine}, we stack all RGS 1 and 2 spectra obtaining two high-quality spectra for both the first and the second order with a total, clean, exposure of 1.529 Ms each. We group the RGS spectra in channels equal to 1/3 of the PSF, and use C-statistics, because it provides the optimal spectral binning and avoids over-sampling. RGS spectral fitting is performed using the SPEX package, with contributions from XSPEC, in particular for reflection models. Flux resolved spectra are extracted using the same good time interval (GTI) files as used for the EPIC-pn analysis.

The \nustar\ data are reduced using the \nustar\ data analysis software (NuSTARDAS) version 1.6.0 and CALDB version 20160731. We extract source counts from a 30$^{\prime\prime}$ diameter circular region, centred on the source, and background counts from a large circular region on the same chip. We combine all the \nustar\ data into a single spectrum, as the count rate is very low due to the extremely soft spectrum, and bin to a signal-to-noise ratio of 6 and oversampling of 3.

\subsection*{Pileup}
In the high flux intervals, the source flux is above the nominal EPIC-pn large window mode pile-up limit of 3~counts~s$^{-1}$\cite{Jethwa15}, hitting $\sim9$~counts~s$^{-1}$ at times. This risks distorting the spectrum and potentially affecting the detection of the UFO. However, the count rate of IRAS~13224-3809 is dominated by photons from below 1~keV (the count rate from 0.5--1~keV is an order of magnitude higher than the 2--3~keV count rate), because it is an extremely soft source. This means that the effects of pile-up are strongest below 2--3~keV. We test this by extracting the same high flux spectrum using an annular region, instead of a circle, with an excised core of $7^{\prime\prime}$ which encircles the central 4 piled up pixels. Above 2~keV, we find no difference in spectral shape between the two spectra, so we conclude that our analysis (restricted to $E>3$~keV) is robust to this effect.
The absorption feature is still present in both the mean spectrum and the low flux spectrum when an annular extraction region is used. We also repeat this test using only single events, and again find no difference. The low flux spectrum and the RGS spectra are not affected by pile-up.

\subsection*{Copper contamination}

\begin{figure}[h]
\centering
\includegraphics[width=0.4\linewidth]{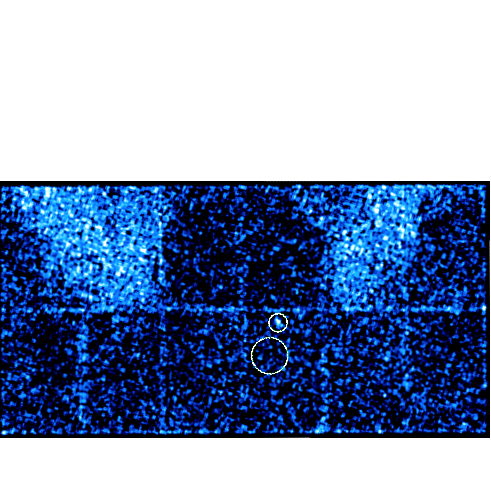}
\includegraphics[width=0.4\linewidth]{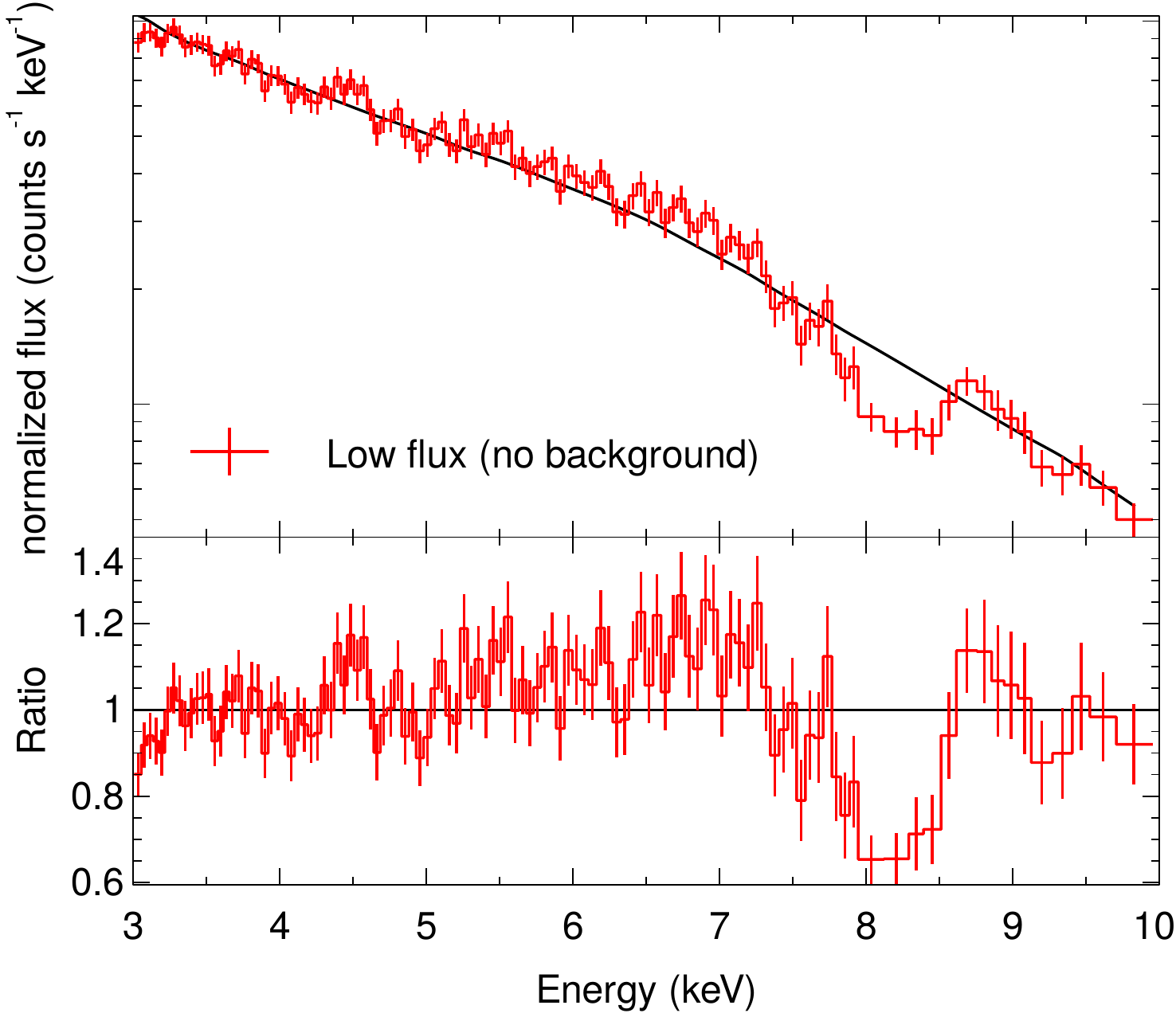}
\caption{Left: 8--8.5~keV EPIC-pn image showing the high instrumental background outside the central chips, with source and background extraction regions marked by white circles, for a representative observation (0780561301). Right: The low-flux spectrum without the background subtracted, fit with a power-law. The UFO line is clearly still visible, and only slightly reduced in strength. All errors are $\pm1\sigma$, and energies are in the observers frame.}
\label{fig_background_map}
\end{figure}

One potential cause of a false detection of a UFO around 8~keV is the complex of emission lines, dominated by Cu K$\alpha$, in the instrumental background\cite{Freyberg04}. Over-subtracting these features would result in an artificial absorption feature at the corresponding energy, which would depend on the source to background flux ratio, giving an anticorrelation between the equivalent width of the line and the source flux. The copper background is only high in the outer regions of the detector, outside the central ~$300^{\prime\prime}$, leaving a central `hole' where contamination is minimal.
We are careful to avoid the region where the copper background is high when selecting background regions, which should prevent contamination (see Extended Data Figure~\ref{fig_background_map}). 
The easiest way to show that the UFO line is not an artefact of background over-subtraction is simply to not subtract the background and check the line remains. While this is not always optimal (it may remove genuine but weak lines, or introduce new features), strong absorption features should remain in the spectrum. In Extended Data Figure~\ref{fig_background_map}, we show the low flux spectrum with no background subtraction, fit with a power-law. The iron line is weaker, due to the additional high energy contribution from the background, but the UFO line is clearly still present.
If the observed line were produced by over-subtraction of the background, the (negative) flux of the line should be constant, the equivalent of an additional constant (positive) line in the background. This is trivial to test, by measuring the strength of an additive line with flux, rather than the multiplicative line we use elsewhere. We find clear variability, and an anti-correlation between the line flux and source flux (Fig.~\ref{fig_linestrength}, inset), which is impossible if the line is a background feature. 
Finally, we note that the lines seen in the RGS spectrum are independent of this effect. We conclude that the line is genuine, and produced by absorption in the AGN spectrum.

The potential secondary feature at $\sim8.7$~keV (observer's frame, 9.2~keV source frame) is coincidental with the Zn~K$\alpha$ line, and appears as an emission feature when the background is not subtracted. We cannot therefore robustly determine whether it is a genuine spectral feature, a statistical fluctuation, or due to the background.

\subsection*{Spectral modelling}

\subsubsection*{EPIC-pn/NuSTAR Stacked Spectra}

We fit the stacked 2016 EPIC-pn spectrum from 3--10~keV (outside the band where pileup effects are present, and where the spectrum is relatively simple and unambiguous), and the stacked \nustar\ spectrum from 3--40~keV. We model the spectrum with the \textsc{relxill} relativistic reflection model \cite{Garcia14}. The relativisitc blurring parameters are consistent with those found by previous authors\cite{Fabian13} (see Table~\ref{table_fitpars}), but a strong absorption feature remains at $\sim8.6$~keV. When we include an additional Gaussian absorption line (modelled with \textsc{gabs}, with $\sigma_\mathrm{Gauss}$ fixed at 0.1~keV), the fit improves by $\Delta\chi^2=26$, for 2 additional free parameters. Parameters for both these models are given in Extended Data Table~\ref{table_fitpars}. We also test allowing $\sigma_\mathrm{Gauss}$ to vary, but we find no significant difference in the fit statistic and no impact on the other fit parameters.

\begin{table}[h]
\centering
\caption{Best fit parameters excluding the absorption feature (model 1) and including it (model 2). Errors are $1\sigma$.}
\label{table_fitpars}
\begin{tabular}{l c c l}
\hline
Parameter & Model 1 & Model 2 & Description/unit\\
\hline
$E_\mathrm{Gauss}$ 	&-& $8.14\pm0.03$ & Line energy (keV)\\
$\sigma_{\mathrm{Gauss}}$	&-& $0.1^*$	& Line width (keV)\\
$N_\mathrm{Gauss}$ & -	& $0.15\pm0.03$ & Line strength\\

$q_\mathrm{in}$	&$>8.5$ &$8.3_{-1.2}^{+0.2}$  & Inner emissivity index\\
$q_\mathrm{out}$&$2.4\pm0.1$ &$2.46\pm0.04$ & Outer emissivity index\\
$r_\mathrm{break}$ & $2.52_{-0.07}^{+0.1}$& $2.59\pm0.04$ & Emissivity break radius $(r_\mathrm{G})$ \\
$a$ & $>0.994$&$0.989\pm0.001$	& Black hole spin\\
$i$ &$60\pm0.1$& $59\pm1$  & Inclination (degrees)\\
$\Gamma$ &$1.99_{-0.06}^{+0.12}$ & $2.06_{-0.09}^{+0.06}$ &Power-law index\\
$E_\mathrm{cut}$  &$50_{-15}^{+17}$ & $50\pm6$ & Power-law cutoff energy (keV)\\
$\log{\xi}$ &$3.25_{-0.09}^{+0.06}$ &$3.34\pm0.01$&Ionization (erg~cm~s$^{-1}$) \\
$A_\mathrm{Fe}$ &$4.4_{-0.5}^{+0.4}$ & $3.5\pm0.2$ & Iron abundance (solar)\\
$R$ &$>9.3$ & $>9.8$ & Reflection fraction\\
$\chi^2/$dof & 314/285 & 288/283 & Fit statistic\\

\hline
\end{tabular}

\vspace{5pt}
Parameters marked with $^*$ are fixed at the stated value.
\end{table}

\begin{figure}[h]
\includegraphics[width=\linewidth]{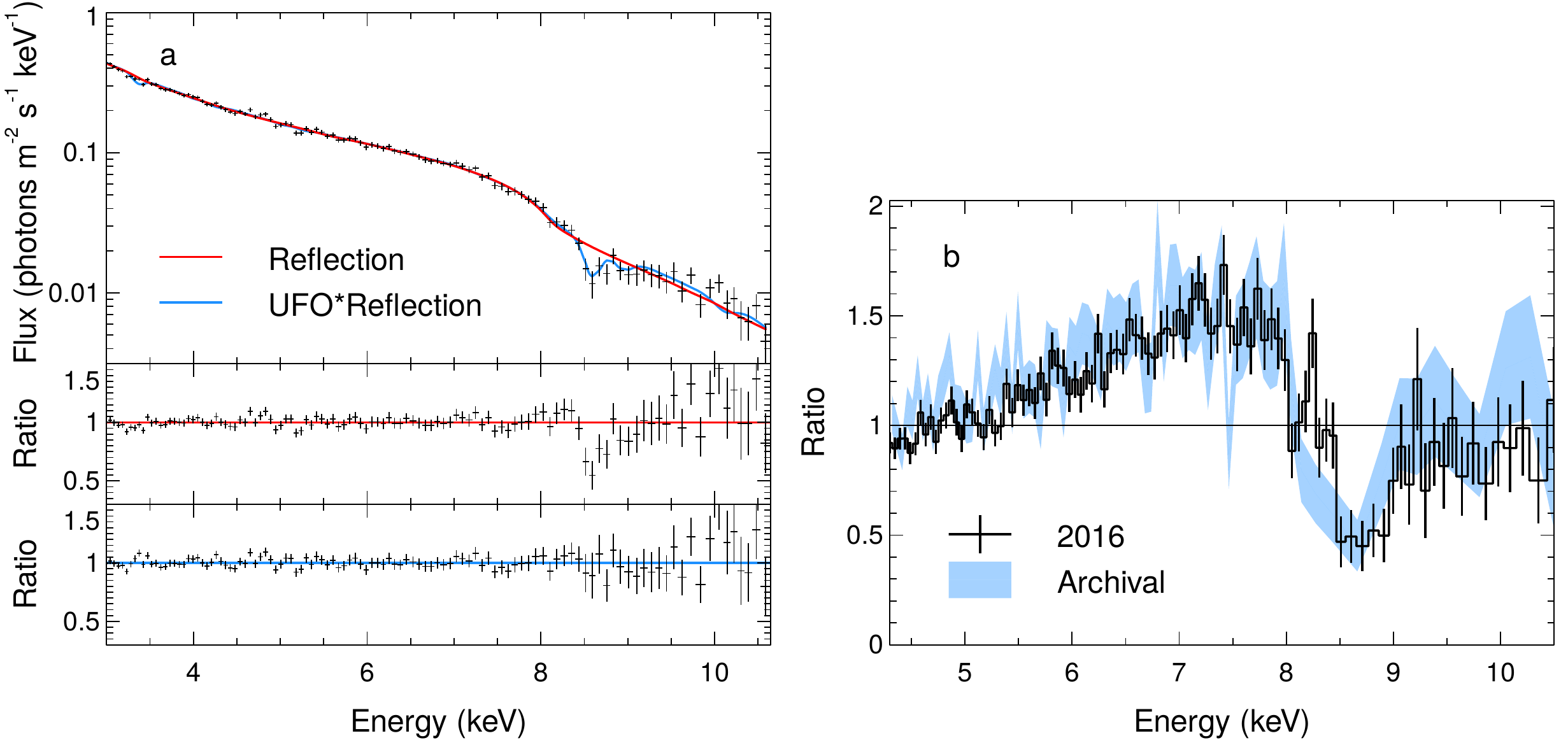}
\caption{\emph{a:} Data and residuals of the EPIC-pn data fit with reflection, with and without an outflowing absorption component. \emph{b:} Residuals of the EPIC-pn data fit with a power-law from 4--5 and 9--10~keV, showing the broad iron line and UFO line. Error bars are $1\sigma$, and energies are in the source rest frame.}
\label{fig_archival_comparison}
\end{figure}

There are some differences between this result and those found by previous authors, which likely stem from the different energy range used. In particular, the photon index, high energy cut-off, and iron abundance are different. The continuum parameters are not of great importance to this work, so long as the continuum is adequately described. The steeper $\Gamma$ in archival results ($\sim2.7$ \cite{Chiang15})  is likely due to the inclusion of the soft excess, which past authors have fit with a two-component reflection model. This requires a steep power-law to produce enough soft photons to fit the soft excess. This model is not unique, as the soft excess generally has limited spectral features due to the lower resolution of the EPIC-pn at these energies, and other factors, such as density of the disk, may alter the parameters from such a fit\cite{Garcia16}. A visual comparison of the archival data and the new data (Extended Data Figure~\ref{fig_archival_comparison} a) does not show any major changes in the structure of the iron line or UFO absorption.

Similarly, the iron abundance is largely determined by the relative strengths of the iron line and soft excess or Compton hump. Given the steep power-law in the dual-reflection model, a high iron abundance is required to produce enough flux in the iron line. This is not required here, as we do not fit the soft excess and the Compton hump is only weakly constrained. This is important, as the iron abundance is potentially degenerate with the strength of the 8.6~keV absorption feature: an increased iron abundance produces a larger iron absorption edge in the reflection spectrum. We can be confident that this is not having a significant effect on our results, because the iron abundance is free to vary in all our fits, including the fits without the absorption modelled, and the feature still remains. We have explicitly searched for degeneracies using MCMC, and find no degeneracy between the strength of the line and the iron abundance.

Following on from this, we performed a blind line scan over the 6--10~keV band, stepping an unresolved Gaussian line ($\sigma_\mathrm{Gauss}=0.01$, allowed to be positive or negative) across the energy band, varying the normalization, and recording the $\Delta\chi^2$ at each point on this grid (Fig.~\ref{fig_fluxresolved}, top). We use the same underlying \textsc{relxill} model, allowing the same parameters to vary. We calculate the significance of this by taking the probability of the maximum $\Delta\chi^2$ for two additional free parameters, and correcting by the number of trials (i.e. the number of resolution elements from 6.7--10~keV). This gives a final chance probability of $1.5\times10^{-5}$, which corresponds to a $4.3\sigma$ detection. No other features are significant above $\sim1\sigma$.

We also fit the absorption with a series of physical models - \textsc{warmabs} in \textsc{Xspec} (shown in Extended Data Figure~\ref{fig_archival_comparison} b), which uses grids of \textsc{xstar} photoionization models, and \textsc{xabs} and \textsc{pion} in \textsc{spex}. The three models give consistent results, with a degeneracy between two possible solutions: $v=0.210\pm0.009$ and $v=0.244\pm0.09$, corresponding to Fe~\textsc{xxv} and Fe~\textsc{xxvi}. These solutions have different column densities and ionizations, which are summarized in Extended Data Table~\ref{table_physicalpn}.
The velocity broadening is not strongly constrained, but does not appear to impact any of the other wind parameters.  We test this by fixing the broadening to lower and higher values, and find no change in the column density, velocity, or ionization of the fit.

\begin{table}
\centering
\caption{Best fit physical model parameters for the stacked EPIC-pn spectrum.}
\label{table_physicalpn}
\begin{tabular}{l c c}
\hline
& Fe~\textsc{xxv} & Fe~\textsc{xxvi}\\
\hline
$v$ ($c$) & $0.244\pm0.09$& $0.210\pm0.009$\\
$\log{\xi}$ (erg~cm~s$^{-1}$) & $4.0\pm0.1$ & $>4.5$\\
$N_\mathrm{H}$ $10^{22}$cm$^{-2}$ & $8.2\pm2.5$ & $180\pm50$\\
$\sigma_{v\mathrm{1D}}$ (km s$^-1$)& $1500\pm1000$ & $1500\pm1000$ \\
\hline

\end{tabular}

\end{table}

\subsubsection*{RGS Stacked Spectrum}

\begin{figure}
\centering
\includegraphics[width=10cm]{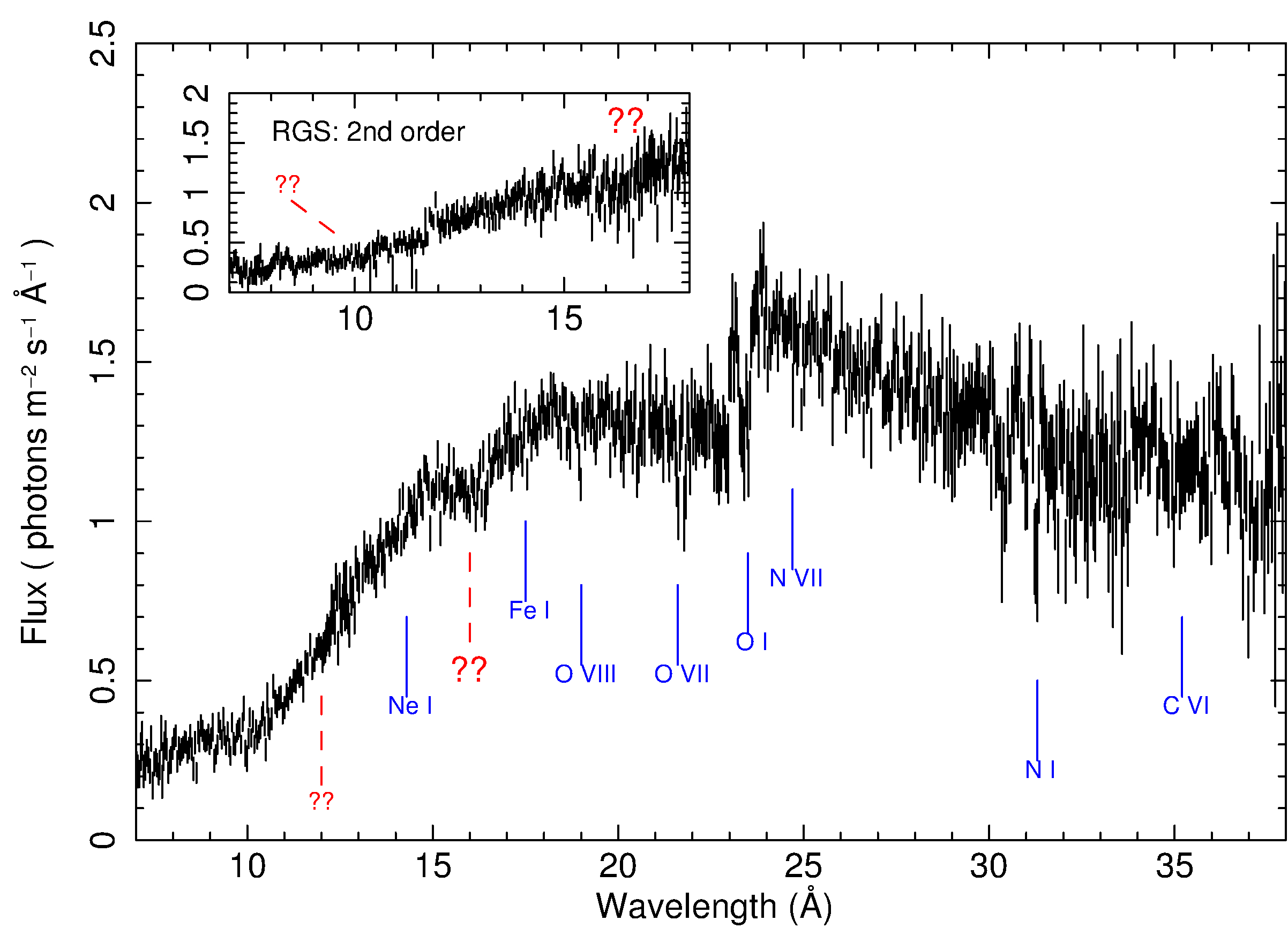}
\caption{RGS first order and second order (small panel) stacked spectra of all 17
observations. Transitions of typical interstellar absorption lines are labelled. Additional features of non-ISM origin are indicated with a dashed red
line and two question marks. Above 33 \AA , the RGS spectrum is affected by high background. Error bars are $1\sigma$, and energies are in the source rest frame.
}
\label{fig_rgs_stacked}
\end{figure}

The RGS spectrum is complex, showing several broad emission-like features at 15~\AA\ and 18~\AA . This spectrometer is the most sensitive to narrow ($\lesssim$1000~\AA ) features, but higher effective area and broader energy range EPIC detectors are more efficient to determine the spectral continuum. We therefore perform an independent analysis of the RGS spectra using either a phenomenological spline continuum model fitted to the RGS spectrum or the physical reflection model provided by the best-fit reflection (relxill) model of the EPIC-pn stacked spectrum. When fitting the RGS spectrum, the spline is corrected for redshift and Galactic interstellar (ISM) absorption. We search for features in the RGS spectrum following an advanced procedure\cite{Pinto16}. We include a Gaussian spanning the 7--38~\AA\ wavelength range in increments of 0.05 \AA , and assume linewidth of 1000~km~s$^{-1}$ (comparable to the RGS resolution). This broadening will also tend to strengthen the detection of any warm-absorber and UFO lines with respect to interstellar absorption lines, since the latter are typically narrower\cite{Pinto13} ($\leq$200~km~s$^{-1}$). We take into account the absorption edges of neutral neon (14.3~\AA ), iron (17.5~\AA ), and oxygen (23.0~\AA ), but we exclude the corresponding 1s--2p absorption lines in order to detect and compare any spectral feature intrinsic to IRAS~13224-3809 or to the ISM. The strongest non-Galactic absorption feature detected is a broad depression around 16~\AA , which is also clear in the RGS stacked spectrum (see Extended Data Figure~\ref{fig_rgs_stacked}). The other two putative, weaker, absorption-like features appear at 10~\AA\ and 13~\AA . Interestingly, the photoionization model of the EPIC spectrum predicts three broad ($\sim$1500~km~s$^{-1}$) UFO absorption lines that match the three RGS absorption features. We have tested different line widths (from 100 to 5000~km~s$^{-1}$) without finding a major effect on their detection. The significance of the rest-frame absorption lines of Galactic O VII and O VIII instead increase up to 5$\sigma$ for narrower widths ($\leq$200~km~s$^{-1}$) confirming the results obtained with the grating spectra of the brightest X-ray binaries\cite{Pinto13}.

A full description of the RGS spectral modelling and the corresponding flux-resolved high-resolution X-ray spectroscopy will be discussed in a forthcoming paper. Here we provide the main result obtained with the overall spectrum and some first interpretation of the wind variability. We have modelled the RGS stacked spectrum with both a spline and a reflection continuum in order to constrain the characteristics of the UFO. The interstellar medium is modelled following the detailed multi-phase gas model constrained with the low-mass X-ray binaries\cite{Pinto13}. We model the UFO absorption features in the RGS spectrum with an outflowing gas in photoionization equilibrium (\textsc{xabs} model in SPEX 3.02). The best fit of the RGS stacked spectrum provides the column density $N_\mathrm{H} = 9.5 \pm 0.5 \times 10^{22}$ cm$^{-2}$ (90\% error) the ionization parameter $\log{\xi}= 4.0 \pm 0.1$~erg~cm~s$^{-1}$ and the line width
$\sigma_v = 2000 \pm 1000$~km~s$^{-1}$. The RGS velocity shift $v = -0.231 \pm 0.007 c$
fits between the EPIC Fe XXV ($-0.244 \pm 0.009 c$) and the Fe XXVI ($-0.210 \pm 0.009 c$) solutions, and does not fully constrain which solution is most likely, but slightly prefers the Fe XXV ($-0.244c$) solution, which is consistent within the 90\% confidence level.

\subsubsection*{Joint fit}

We investigate a combined fit to both EPIC-pn and RGS spectra, fitting with the same absorption model but different continuum models for each spectrum (the physical reflection model for the pn, and a spline for the RGS). We also include a photoionized emission component, modelled with \textsc{photemis}. The soft and hard absorption features are consistent with being from the same absorber (freeing the parameters between the two results in an improvement to the fit of only $\Delta\chi^2=3$, for 4 additional free parameters). 
The joint fit clearly prefers the Fe~\textsc{xxv} solution, with final best fit parameters of $v=0.236\pm0.006c$, $\sigma_v=4000\pm1000$, $\log(\xi)=4.14\pm0.13$ and $N_\mathrm{H}=2.2^{+0.8}_{-1.6}\times10^{23}$~cm$^{-2}$. The increased broadening with respect to the individual spectrum fits may be due to a small offset between the EPIC-pn and RGS spectra, which could be caused by gain shift in the EPIC-pn. However, it is consistent at the 90\% level with that found from the RGS alone.

The inclusion of the emission component improves the fit significantly ($\Delta\chi^2=21$, for 2 additional free parameters), accounting for the residuals at $\sim8.3$~keV and other possible features. The velocity of this component is $0.213\pm0.015c$, and the luminosity $1.1\pm0.5\times10^{41}$ erg~s$^{-1}$. If this component is genuine, it is made up of scattered emission from the wind, and can in principle be used to determine the wind geometry. However, it is likely that much of the P-Cygni profile, including any redshifted emission, is obscured by the relativistic iron line, which is very strong in this source. One possible approach to take here would be searching for the emission component of the P-Cygni profile in the lag spectra, as the reverberation timescale should be much longer than for the relativistic reflection component, due to the greater distance from the source.

\subsubsection*{EPIC-pn Flux-resolved Spectra}
We also fit the three flux-resolved spectra, tying parameters we expect to be constant (such as $a$ and $i$) between the different spectra. The model parameters are consistent with those given in Table~\ref{table_fitpars}, with the reflection fraction inversely proportional to flux. We perform the same line scan over these spectra simultaneously, stepping the line across in energy then recording the $\Delta\chi^2$ for each spectrum individually. The line is only significantly detected in the low flux spectrum, with a maximum $\Delta\chi^2$ of 59.7, for 2 additional free parameters. This gives a corrected probability of $1.96\times10^{-12}$, and a significance of $7.0\sigma$. 

We also check the robustness of the low-flux line detection using a Monte-Carlo test. We draw parameters from an MCMC chain, used to evaluate the errors and degeneracies in the best fit parameters, and use them to simulate 10,000 fake spectra. We then fit these spectra using the same procedure. None of the simulated spectra have higher significance features, in either emission or absorption, setting a lower limit of $P>99.99\%$ on the significance. Given the expected fraction of $4.6\times10^{-12}$, it is unfeasible to test sufficient spectra to establish the true significance with this method.

We perform a similar analysis with 10 flux-resolved spectra, again with the same number of counts in each. We fit the spectra simultaneously, using \textsc{relxill} and a Gaussian absorption line, allowing the reflection fraction, power-law index, and normalizations of the reflection and Gaussian components to vary between each spectrum. This gives a reasonable fit ($\chi^2/$d.o.f. $=966/857=1.13$). We then record the equivalent width and flux of the absorption line in each spectrum, and the 3--10~keV flux. These are plotted against each other in Fig.~\ref{fig_linestrength}, showing a strong correlation. We use Bayesian regression to perform a linear fit which incorporates the upper limits, and draw samples from the posterior distribution to calculate the uncertainty. We calculate the probability of a stronger correlation being found from a constant absorption feature by simulating 10,000 sets of points with the same errors, assuming that the line strength is constant in each case, and performing the same analysis. In no case do we find a stronger correlation.

\subsubsection*{RGS Flux-resolved Spectra}
We perform high-resolution flux-resolved X-ray spectroscopy with the RGS data, consistent with that performed with EPIC: we extract RGS 1 and 2 first- and second-order spectra with the good time intervals defined according to the EPIC flux-prescriptions. We stack the RGS 1 and 2 spectra for each flux-range obtaining three high-quality RGS spectra with comparable statistics. 
As previously seen for the overall stacked spectrum, there are some non-interstellar absorption-like features (9.5~\AA , 13~\AA , 16~\AA ) which show evidence of variability, being both stronger and possibly bluer in the low-flux spectrum. In order to probe the strength of the features in each spectrum, we apply the same technique used for the stacked spectrum by fitting a Gaussian over the 7--38~\AA\ wavelength range in increments of 0.05~\AA . In Fig.~\ref{fig_rgs_significance}, we show the significance of the spectral features obtained adopting the (RGS fitted) spline continuum. The three broad features are still detected at 9.5, 13.0, and 16~\AA\ in the low-flux spectrum. They have less significance or are undetected in the higher flux spectra with possible evidence of slight velocity change. Their wavelengths match with the strongest lines predicted by the 0.24c UFO model in the RGS energy band: $10.0 \pm 0.5$~\AA\ (Ne X + Fe XVIII-XXII blend), $13.2 \pm 0.5$~\AA\ (O VIII K$\beta$ + Fe XVIII) and $15.8 \pm 0.5$ (O VIII K$\alpha$). The strength of the absorption lines anti-correlate with the flux in agreement with the EPIC result and therefore strongly argues in favour of a connection between the EPIC and RGS absorbers as being part of the same extreme wind. We computed the confidence level of the three main absorption lines in the low-flux spectrum, where they are significantly detected as in EPIC. Accounting for the number of trials due to bins of 0.05~\AA\ and an outflow-velocity range from 0-to-0.3c, we obtain 2.1$\sigma$, 2.9$\sigma$, and 3.4$\sigma$ for the 9.5~\AA , 13~\AA , and 16~\AA\ absorption lines, respectively, which -- since they have the same velocity shift -- gives a cumulative 5.1$\sigma$ detection.


\subsection*{Energetics}

We can estimate the mass outflow rate by combining the velocity and column densities\cite{Nardini15}, and the mass of $6\times10^{6}M_\odot$\cite{Zhou05} (estimated using the empirical reverberation relation\cite{Kaspi00}):
\begin{equation}
\dot{M}=\Omega N_\mathrm{H} m_\mathrm{p} v R_\mathrm{wind}
\end{equation}
where $\Omega$ is the solid angle of the wind and $R_\mathrm{wind}$ the radius.

We cannot be confident of the value of $\Omega$, as the emission from the P-Cygni profile, if there is any, is obscured by the blue horn of the iron line, which is extremely strong in this source. However, given that the absorption line is found in the stacked archival data (most of which is from 2011), this implies that the feature has been present and roughly constant (as a function of flux) for at least 5 years, which would argue for a reasonably large covering fraction, otherwise any clump along the line of sight would likely have moved away. 
Similarly, we do not know the radius of the wind. However, we know that it must be variable on timescales $\lesssim5$~ks, which corresponds to 170~$R_\mathrm{G}$. 
Assuming a radius of 100~$R_\mathrm{G}$, we find $\dot{M}=2\times 10^{23}\Omega$~g~s$^{-1}$ ($0.03\Omega\ M_\odot$~year$^{-1}$) for Fe~\textsc{xxv}, while the Eddington accretion rate for a black hole of this mass is $2.7\times 10^{24}$~g~s$^{-1}$, assuming an efficiency of 0.3 for near-maximal spin. In either case, a large fraction of the matter accreted by the disk is lost to the wind, possibly implying super-Eddington accretion at large radii.

We can then calculate the power in the wind:
\begin{equation}
\begin{split}
P &= \frac{1}{2}\dot{M} v^2 \\
P &= 0.006\Omega L_\mathrm{Edd}\\
\end{split}
\end{equation}

For $\Omega=2\pi$, this gives a power of 4\% of the Eddington luminosity, implying that a significant fraction of the accretion power must be lost into the wind. For the same assumed covering fraction, the power of PDS~456 is 15\% of the Eddington luminosity\cite{Nardini15}.

\subsection*{Alternative Interpretations}
The prevailing interpretation for highly blueshifted absorption features in the X-ray spectra of AGN is that they are due to outflowing gas. However, it is possible that some of these features may instead be due to absorption by a diffuse absorbing surface layer on the approaching side of the accretion disc, which naturally gives relativistic velocities \cite{Gallo13}.
The absorption line then appears in the reflection component. Aberration means that the blue side is brighter than the red side. For a disc inclination of 60 degrees the absorption layer needs to extend from $\sim5$ to 10~$R_\mathrm{G}$  to give an observed  line at 8.2~keV. If the brighter parts of the light curve are associated with the corona rising above 10~$R_\mathrm{G}$, then reduced light bending and irradiation of the inner disc weakens both the reflection component and the absorption, consistent with observation. 

It is also possible that the variability is produced by a geometric effect. Previous authors have suggested that the relatively constant spectrum of the relativistic reflection component can be produced by changes in the height or extent of the X-ray corona above the disk\cite{Miniutti04}, and the covering fraction of a low scale-height wind could similarly depend on the size or position of the compact X-ray source.

\subsection*{Code availability}
All the code used for the data reduction is available from the respective websites. \textsc{xspec} and \textsc{spex} are freely available online. Code used for generating figures, calculating flux-resolved extraction intervals, and calculating line significance, is available upon request to the lead author.

\end{document}